\documentclass[12pt]{article}
\usepackage[italian,english]{babel}
\usepackage{graphicx}
\usepackage{psfig,picinpar,floatflt,epsf}
\usepackage{latexsym}
\usepackage{amsmath,amsfonts,amssymb,amsthm,wasysym}
\def\bseq{\begin{subequation}}  
\def\eseq{\end{subequation}}
\def\bsea{\begin{subeqnarray}}  
\def\esea{\end{subeqnarray}}
\def\Bar#1{\overline{#1}}                       

\newcommand{\bbox}{\lower.2ex\hbox{$\Box$}}

\newcommand{\beq}{\begin{equation}}
\newcommand{\eeq}{\end{equation}}
\newcommand{\bea}{\begin{eqnarray}}
\newcommand{\eea}{\end{eqnarray}}
\newcommand{\ena}{\end{eqnarray}}

\newcommand {\non}{\nonumber}

\renewcommand{\a}{\alpha}

\renewcommand{\d}{\delta}
\newcommand{\pa}{\partial}

\newcommand{\G}{\Gamma}

\newcommand{\D}{\Delta}

\newcommand{\mb}{\overline{m}}

\newcommand{\F}{\Phi}

\newcommand{\Fib}{\bar{\Phi}}

\newcommand{\s}{\sigma}

\newcommand{\te}{\theta}
\newcommand{\teb}{\bar{\theta}}

\newcommand{\Tr}{{\rm Tr}}
\renewcommand{\(}{\left(}
\renewcommand{\)}{\right)}
\renewcommand{\[}{\left[}
\renewcommand{\]}{\right]}
\newcommand{\ggl}{\left\{}
\newcommand{\ggr}{\right\}}

\newcommand{\Db}{\bar{D}}

\newcommand{\CN}{{\cal N}}

\newcommand{\be}{\begin{equation}}
\newcommand{\ee}{\end{equation}}
\newcommand{\mc}{\mathcal}

\begin{document}
\begin{titlepage}
{\hbox to\hsize{January 2006 \hfill}}
\begin{center}
\vglue .06in
\vskip 40pt

{\Large\bf On supersymmetry breaking and} \\ 
[.07in]
{\Large\bf the Dijkgraaf-Vafa conjecture} 
\\[.7in]

{\large\bf L. Girardello\footnote{luciano.girardello@mib.infn.it},
A. Mariotti\footnote{alberto.mariotti@mib.infn.it}
and G. Tartaglino-Mazzucchelli\footnote{gabriele.tartaglino@mib.infn.it}}%
\\[.5in]
{\it Dipartimento di Fisica, Universit\`a degli studi di
Milano-Bicocca\\ 
and INFN, Sezione di Milano-Bicocca, piazza delle Scienze 3, I-20126 Milano, 
Italy}\\[.7in]

{\bf ABSTRACT}\\[.3in]
\end{center}
We investigate the Dijkgraaf-Vafa proposal when supersymmetry is broken.
We consider $U(N)$ SYM with chiral adjoint matter where the coupling constants 
in the tree-level superpotential are promoted to chiral spurions.
The holomorphic part of the low-energy glueball superpotential can still be 
analyzed. We compute the holomorphic supersymmetry breaking contributions
using methods of the geometry underlying the $\CN=1$ effective gauge theory
viewed as a Whitham system. 
We also study the change in the effective glueball superpotential
 using perturbative supergraph techniques in the presence of spurions.

\vskip 40pt
${~~~}$ \newline

\end{titlepage}

\section{Introduction}

Dijkgraaf and Vafa have proposed that the low-energy glueball effective 
superpotential of $\mc{N}=1$ supersymmetric gauge theories in four dimensions 
can be computed via an auxiliary matrix model \cite{DV}.
The simplest case is a $U(N)$ gauge theory coupled to a massive adjoint chiral 
matter multiplet $\F$
with a tree--level superpotential $W(\Phi)$.
The proposal stems from a set of string dualities in the framework of
geometrically engineered gauge theories, 
topological strings and matrix models \cite{LargeN,cachazovafa,DV}.
The large-N matrix model analysis brings in an algebraic curve which may
correspond to a Calabi--Yau dual geometry \cite{LargeN}. We shall consider 
gauge theories 
that can be obtained from string theories that lead to such geometries.
The DV proposal has been tested and supported directly on the field
theoretical side by perturbative computation
via superfields formalism \cite{DGLVZ} and then by using arguments based on
anomaly equations \cite{CDSW}.

We study here the case where susy is broken explicitly (soft and/or non soft) 
by the introduction of 
spurionic fields \cite{GirardelloGrisaru}.
Holomorphy at large is lost, but holomorphic
quantities such as the glueball superpotential can be still analyzed and one
can compare the computation in the superfields formalism adapted to
spurion fields with that one using
the algebraic curve underlying the effective gauge theory.
In order to discuss such breaking
we utilize two notions:
a closed string realization of the method
of the spurions \cite{LawMcG} and Whitham deformations 
\cite{Gorsky,cekov}.

It is worth to recall the geometrical origin of such gauge
theories for type $IIB$ string theory  in order to insert
the notion of spurion in a natural
way in this language.
We have in mind $D$-branes partially wrapped over non trivial
$2$-cycles of non compact CY and the dual description where
$D$-branes have been replaced by fluxes \cite{LargeN}.
In the UV, adjoint chiral multiplets $\Phi$ arise from holomorphic deformations
of the supersymmetric cycles and of open string gauge bundles on these cycles.
A four-dimensional superpotential for these fields 
can arise and can be written as $W=W(\Phi,g_k)$
where $g_k$ depend only on the complex structure.
From the perspective of the $D3$--brane action
in the low--energy limit, where supergravity decouples,
the $g_k$ can be interpreted as couplings.
As already suggested in \cite{LawMcG}, 
the susy breaking parameters are described by auxiliary
components of the closed string fields, typically magnetic
fluxes along CY directions, depending on the complex
structure moduli. Such fluxes are introduced by hand
without back reaction of the string or of the supergravity
backgrounds.
In the four dimensional supergravity language they are $F$--components
of chiral multiplets which depend only on the complex structure moduli.
Vev of such $F$--terms cause spontaneous breaking of local susy
and, in the appropriate flat limit with decoupling of
supergravity, they appear as explicit breaking terms
which can be written in the
spurionic fashion in the rigid susy action.

A non--perturbative analysis of susy broken effective dynamics
has been done in \cite{Marino} for $\mc{N}=2$ supersymmetric gauge theories.
In that context the connection between the Seiberg--Witten solution \cite{SW}
and integrable systems (Whitham hierarchy) \cite{Gorsky} was 
used.
The authors of \cite{Marino} break susy promoting the Whitham parameters of the
hierarchy to spurions and then compute the broken effective potential
using the $\CN=2$ integrable structure.\\
As in the $\CN=2$ case, a relation between the Whitham systems and the $\CN=1$ 
effective geometry was established in \cite{cekov}.
This suggests to break supersymmetry promoting the Whitham parameters to 
spurions as in the $\CN=2$ case.
In the $\CN=1$ geometry the Whitham parameters 
are precisely the tree--level coupling costants of the
matter superpotential \cite{cekov}.
We will break the $\CN=1$ supersymmetry promoting 
them to spurions, and 
the Whitham hierarchy can then be interpreted as a family of
supersymmetry breaking deformations of the original theory. 
Using this interpretation, we will compute directly from the geometrical data
the holomorphic supersymmetry breaking contributions in the 
low--energy effective glueball superpotential.

We have also analyzed with perturbative supergraph techniques the effective 
glueball superpotential when susy is broken with spurions. Arguments for the 
computability of the effective superpotential have been presented in 
\cite{LawMcG}. If supersymmetry
is broken, holomorphicity in the coupling constants is no longer guaranteed,
the computation is much harder than in the $\CN=1$ case and the
simplifications of \cite{DGLVZ} do not work in general. 
Anyway, we can restrict ourselves to a particular subclass of contributions 
for which a spurionic superfields 
generalization of the techniques in \cite{DGLVZ} can be done.
Within such strong approximation and with unbroken $U(N)$ gauge group,
we find that to all order in the glueball superfield the effective 
superpotential 
has the same functional form of the $\CN=1$ case where
the coupling constants are replaced by spurions and so it results still 
holomorphic.

The paper is organized as follows: In section 2 we review the geometry 
underlying the Dijkgraaf--Vafa proposal. 
In section 3 we introduce 
supersymmetry breaking by spurions and discuss 
the low--energy glueball superpotential.
In section 4 we discuss the geometry as 
a Whitham system and use it in the susy broken case.
In section 5 we treat the explicit example of a deformed susy broken cubic 
tree--level superpotential.
In section 6 we use perturbative superspace techniques 
along the line mentioned above. 
Section 7 is devoted to conclusions. At the end, in two appendices, 
we describe the computational details of section 5 and section 6.

\section{The geometrical picture}\label{DVreview}

We consider the particular case of a $\mc{N}=1$, $U(N)$ gauge
theory with a degree $n+1$ polynomial tree--level superpotential
$W(\Phi)$ for the chiral matter superfields in the adjoint
representation of the gauge group
\be
W(\Phi)=\sum_{k=1}^{n+1}\frac{g_k}{k} \Tr\,\Phi^k~.
\label{Wtree}
\ee
In a generic vacuum the gauge group $U(N)$ is broken to
$U(N_1) \times \dots \times  U(N_n)$. 
In the IR limit 
the effective low--energy degrees of freedom are described 
by the glueball superfields 
$S_i={1\over 32\pi^2}\Tr\,W^{\alpha}_i W_{\alpha i}$ where 
 $W^{\alpha}_i$ is the fermionic chiral superfield, field strength
of the vector multiplet of the unbroken gauge group $U(N_i)$.

The expression for the non perturbative glueball superpotential reads
\be
W_{eff}(S_i)=-\sum_{i=1}^n \[N_i\frac{\pa \mc{F}}{\pa S_i}+2\pi i
\tau_{i}S_i\]~,
\ee
where $\mc{F}$ is the prepotential which can be computed from the geometrical 
data \cite{LargeN,cachazovafa}. 
In \cite{DV} it has been proposed to reinterpret and compute this prepotential
as the free energy of an associated matrix model. In \cite{CDSW,witten}
it was also deduced directly on the field theoretical ground 
using generalized Konishi anomaly equations.

The geometry associated with the low--energy theory is described 
by a family of genus $g=n-1$ Riemann surfaces and by a meromorphic differential
$dS$
\bea
\label{curve}
y^2&=&[W(x)']^2+f^{(n-1)}(x)~,\\
\label{diffDV}
dS&=&y \, dx =\sqrt{[W(x)']^2+f^{(n-1)}(x)} \, dx~.
\eea
The degree $n-1$ polynomial 
$
f^{(n-1)}(x)=\sum_{l=0}^{n-1}f_l x^l~,
\label{fModuli}
$
is associated with the quantum contributions and  the coefficients $f_l$
($l=0,\cdots,n-2$)
are the moduli of the complex curve; the derivatives of the
meromorphic differential (\ref{diffDV}) with respect to the moduli
gives holomorphic differentials.

A basis of canonical cycles 
\cite{cekov,witten} 
is  $\{\alpha^i,\beta_i,\alpha^0,\beta_0 \}$, where
$i=2,\dots,n$, with intersection numbers
$(\beta_b  \cap \alpha^a=\delta^a_b)$. The cycles are all compact except
$\beta_0$.
We label the cuts starting from the larger real root of the algebraic curve
(\ref{curve}), so from right to left.
The 
$\alpha^i$--cycle surrounds counterclockwise
the $i$--th cut
while the $\alpha^0$--cycle encircles all the cuts and
then gives the residue at infinity. The dual $\beta_i$--cycle
($i=2,\dots,n$) passes clockwise through the $i$-th and the first cut, while 
$\beta_0$ goes from the second sheet infinity to the first passing through the 
first cut. The periods $s_i$, the parameter $t_0$ and the conjugated periods
are defined as
\bea
s_i=\oint_{\alpha^i} dS~, 
&\qquad&
t_0=\oint_{\alpha^0} dS=-Res_{\infty}(dS)=\frac{f_{n-1}}{2 g_{n+1}}~,
\label{SiPi1}
\\
\Pi_i=\frac{1}{2}\oint_{\beta_i} dS~,
&\qquad&
\Pi_0=\frac{1}{2} \int_{\beta_0} dS~.
\label{SiPi2}
\eea
In these variables the effective superpotential computed by the geometry is
\be
-W_{eff}=N \Pi_0+\sum_{i=2}^{n} N_i
\Pi_i= N \frac{\pa \mc{F}}{\pa t_0}+\sum_{i=2}^{n} N_i \frac{\pa
\mc{F}}{\pa s_i}~~,
\label{WeffS}
\ee
where $\sum_{j=1}^{n} N_j=N$.
In the previous formula we
have introduced the prepotential\footnote{The prepotential differs
from the usual one \cite{DV,LargeN} for a multiplicative factor due to the
change of variables. Anyway, this difference is not felt by the
effective superpotential $W_{eff}$ which is a function of the dual
periods, the quantities which really enter in the computation, as
in (\ref{WeffS}).} $\mc{F}$ such that its derivatives w.r.t. the
$\{s_i, t_0\}$ periods give the dual ones $\{\Pi_i, \Pi_0\}$.

Upon getting the superpotential as a function of the variables $s_i$ 
and $t_0$, we return to the variables of \cite{DV,LargeN} 
using\footnote{The variables of  \cite{DV,LargeN} are 
$S_j=-\frac{1}{2}\oint_{A^j} dS$ and 
$\Pi_j=\frac{1}{2}\oint_{B_j} dS$,
with $\{A_j,B^j;~j=1,\dots,n\}$ a different set of cycles with
all $B_j$ non compact.}
\bea
s_i&=&-2 S_i~~,~~~~~\,~~~~~~~~ i=2,\dots,n~~,\non\\
t_0&=&-2\sum_{j=1}^{n}S_j~~,
\label{cambcoord}
\eea
in fact the $S_i$ are the physical variables which are
interpreted as the glueball superfields.

\section{Supersymmetry breaking}

The introduction of spurionic fields provides the standard mechanism for the 
explicit (soft and/or non soft) breaking of global supersymmetry.
In the $\CN=1$ case the tree--level superpotential $W_{tree}$, and
the effective glueball prepotential $\mc{F}$, depend on the coupling constants 
$g_m$ associated with the operators $\Tr\,\Phi^m$ in the ultraviolet action. 
In order to break $\mc{N}=1$ supersymmetry down to $\mc{N}=0$ we promote the 
coupling constants $g_m$ to $\mc{N}=1$ chiral superfields $G_m$ and then we 
freeze
the scalar and the auxiliary $F$--components to constant values. 
In this way the chiral spurions $G_m=g_m+\te^2\Gamma_m$ produce non 
supersymmetric terms in the
superpotential $W_{tree}$.
We want to study their effects on the low energy glueball
effective superpotential under the assumption that the low 
energy degrees of 
freedom are still the glueballs. 
The breaking parameters $\Gamma_m$ must be considered the 
smallest scales in the theory. They are thought as
small perturbations of the $\CN=1$ theory by keeping fixed 
the $\CN=1$ vacuum structure and the gauge symmetry breaking patterns
$U(N)\to U(N_1)\times\cdots\times U(N_n)$.

We set the scalar components of $G_m$ equal to the coupling constants 
$g_m$ for $m\leq n+1$, 
zero for $m>n+1$, and the $F$--components $\Gamma_{m}$ will be considered as 
small susy breaking parameters for all $G_m$. Explicitly
\bea
\label{spurios1}
G_k&=& g_k+\theta^2 \Gamma_k ~~~\,,~~k \leq n+1~~, \\
\label{spurios2}
G_j&=& \theta^2 \Gamma_j ~~~~~~~~~~,~~  j > n+1~~,
\eea
and hence we will consider tree--level superpotential (\ref{Wtree})
perturbed as
\be
\label{Wtreedef} 
W_{tree}(\Phi)=\sum_{k=1}^{n+1}\frac{G_k}{k} \Tr\,\Phi^k +\theta^2 
\sum_{j>n+1}
\frac{\Gamma_{j}}{j} \Tr\,\Phi^j~~.
\ee
Notice that besides having promoted to spurion the coupling constants
already appearing in the tree--level superpotential,
we have also added pure auxiliary $F$--terms. For $k>3$ these spurionic terms 
are
not soft and quadratic divergences can appear in the wave function 
renormalization; in any case they have to be considered as dangerously 
irrelevant operators with the usual warning \cite{Kutasov,LawMcG}.
The $\Gamma_m$ for $m \leq n+1$ can be interpreted as vacuum 
expectations values of 
fluxes \cite{LawMcG}, whereas it is not obvious that this 
is the case for $m > n+1$.
In any case, we will see that the generalization to all the $\Gamma_m$ terms 
is of some interest in the application of the Witham approach.

Let now analize what happens in the effective theory when the
$\G_m$ are turned on. We will restrict ourselves to a discussion of some formal
aspects 
which can be extracted from the geometry of the $\CN=1$ case.
We assume that in the effective dynamics the emergence of the spurions $G_m$ 
are controlled by the holomorphic dependence of the $\CN=1$ prepotential 
$\mc{F}(S_i,g_m)$ on the coupling constants. If we restrict ourselves to 
holomorphic terms in the low--energy glueball superpotential, the prepotential 
in the susy broken phase has the same functional form as the $\CN=1$ case where
now the coupling constants $g_m$ are replaced by the spurions as $G_m$.
This is essentially a naturalness assumption on the effective
superpotential \cite{Seiberg9309335}. 
In section 6 we will discuss 
these assumptions using superfields perturbative techniques
extending \cite{DGLVZ} to the susy broken case.\\
We make some comments about the interpretation
of the couplings $\Gamma_j$ ($j>n+1$). They must be 
understood as coming from tree--level superpotential $W_{tree}$
 of degree greater than
$n+1$ where also the scalar coupling constants $g_j$ above the 
$(n+1)$--degree are turned on. The low energy glueball prepotential
will also depend on all these couplings. We then 
consider the effective theory of (\ref{Wtreedef}) as obtained from that one of 
higher degree in the limit where 
$g_j\to 0$ ($j>n+1$) and in the same vacuum of the theory of 
($n+1$)--degree\footnote{We can choose $\CN=1$ massive theories 
with classical vacua
configuration which are
 nonsingular for $g_j \to 0$ such that there is analitycity of
the glueball 
superpotential around $g_j=0$.}.
In conclusion the prepotential depends on the $n$ glueball superfields $S_i$ 
(in our conventions $t_0$ and $s_i$) and it is evaluated where $g_j=0$.

We expand now the prepotential $\mc{F}(S_i,G_m)$ around the 
supersymmetric vacuum.
If we consider the case of broken supersymmetry with 
$G_m$ having the form (\ref{spurios1}, \ref{spurios2}) 
the terms with more than one power of $\Gamma_m$ will not give any 
contribution and we have
\bea
\mc{F}(s_i,g_k,\Gamma_k,\Gamma_j)&=&\mc{F}(s_i,g_m)|_{g_j=0}~+\non\\
&+&\theta^2~\sum_{k=1}^{n+1}\Gamma_k 
{\frac{\pa\mc{F}(s_i,g_m)}{\pa g_k}}\Big{|}_{g_j=0} ~+~
\theta^2\sum_{j>n+1}\Gamma_{j}{\frac{\pa\mc{F}(s_i,g_m)}{\pa g_j}}
\Big{|}_{g_j=0}~~.~~~
\eea
The first term in this expression is the prepotential of the 
supersymmetric case for a theory with tree--level superpotential of 
$(n+1)$--degree. As just discussed the last term is interpreted
as coming from an higher degree theory in the appropriate limit.\\
We now insert this expression in (\ref{WeffS}) and we obtain the 
holomorphic glueball
superpotential associated with a tree--level susy breaking superpotential as 
(\ref{Wtreedef})
\bea
-W_{eff}&=&N\Bigg{[}\frac{\pa \mc{F}}{\pa t_0}+
\theta^2~\sum_{k=1}^{n+1}\Gamma_k
\frac{\pa^2 \mc{F}}{\pa t_0 \pa g_k}+\theta^2 
\sum_{j>n+1}\Gamma_j\frac{\pa^2 \mc{F}}{\pa t_0 \pa g_j}\Bigg{]}+\non\\
&+&\sum_{i=2}^{n}N_i\Bigg{[} \frac{\pa \mc{F}}{\pa
s_i}+\theta^2~\sum_{k=1}^{n+1}\Gamma_k \frac{\pa^2 \mc{F}}{\pa s_i
\pa g_k}+\theta^2
\sum_{j>n+1}\Gamma_j \frac{\pa^2 \mc{F}}{\pa
s_i \pa g_j}\Bigg{]}~.
\label{223}
\eea
In this expression the first terms within the square bracket are 
supersymmetric, 
whereas the others break susy explicitily: they involve second derivatives of
the prepotential evaluated where the $g_j\equiv 0$ ($j>n+1$).
We will show in the next section how to obtain directly and efficiently from 
the 
geometrical data of the $\CN=1$ theory 
the mixed second derivatives of $\mc{F}$ appearing in (\ref{223}) in order to 
extract the effective supersymmetry breaking contributions.

\section{The $\CN=1$ geometry and Whitham systems}

The geometry of the $\CN=1$ low--energy effective theory is associated with the
generating meromorphic 
differential $dS$ (\ref{diffDV}) and it can be thought as coming from a 
Seiberg--Witten geometry of a $\mc{N}=2$ theory \cite{cachazovafa}.
The addition of a superpotential together with a geometric transition
and a desingularization leads to such geometry
with parameters $g_k$ and complex moduli $f_l$ \cite{LargeN}.
The couplings $g_k$ can be viewed as Whitham
deformations of the previous SW geometry.
Performing a Whitham deformation mean extending the
parameter space of the curve with extra variables \cite{Gorsky}.
As a consequence of this deformation the moduli of the curve 
and also the generating differential
become functions of these new parameters.

As shown in \cite{cekov} the $\CN=1$ geometry can be
embedded into the Whitham framework.
The moduli $f_l$ of the curve (\ref{curve}) are 
functions $f_l=f_l(g_k,t_0,s_i)$ of the Whitham parameters $g_k$ and of 
$(t_0,\,s_i)$, the
periods of the generating differential $dS$ along the $\alpha$--cycles. 
We review some results of \cite{cekov} and set up our conventions.\\
One of the advantages we gain using Whitham description is that it
provides an efficient way to compute the mixed second derivatives 
appearing in (\ref{223}) directly in terms of geometrical data since the 
coupling constants are considered as independent parameters.

Using the whole set of variables ($g_k,\,t_0,\,s_i$) characterizing the curve
(\ref{curve}) and the generating differential (\ref{diffDV}), the Whitham
system can be defined by the following set of equations \cite{cekov}
\be
\frac{\partial dS}{\partial s_i}=d\omega_i~~,
\qquad \frac{\partial dS}{\partial t_0}=d\Omega_0~~,
\qquad \frac{\partial dS}{\partial g_k}=d\Omega_k~~,
\label{dIf}
\ee
where $d\omega_i$ are normalized holomorphic differentials
\be
\oint_{\alpha_i} \frac{\partial dS}{\partial s_j}=
\oint_{\alpha_i} d\omega_j=\delta_{ij}~~.
\ee
The differentials $d\Omega_k$ are meromorphic of the second kind with poles 
only at the infinity points $\pm \infty$;
$d\Omega_0$ is a differential of the third kind with residue at
$\pm \infty$. 
They have vanishing $\alpha$--periods and behave at infinity 
as ($\xi=\frac{1}{x}$)
\be
\label{normcond}
\oint_{\alpha^i}d\Omega_0=\frac{\pa s_i}{\pa t_0}=0~~, 
\quad \oint_{\alpha^i}d\Omega_k=\frac{\pa s_i}{\pa g_k}=0~~;
\qquad d\Omega_l=-(\xi^{-l-1}+O(1))d\xi~~.
\ee
These normalization conditions characterize $s_i,t_0$ and $g_k$ as independent
variables.\\
The generating differential $dS$ is then
a linear combination of the differentials 
(\ref{dIf})
\beq 
\label{diffgeneds}
dS=\sum_{i=2}^{n} s_i \, d\omega_i+t_0 \,
d\Omega_0+\sum_{k=1}^{n+1} g_k \, d\Omega_k~= \sqrt{[W(x)']^2+
\sum_{k=0}^{n-2}f_{k}x^k+2g_{n+1} t_0 x^{n-1}} \, dx~~.
\eeq 
Consinstency of the equality in (\ref{diffgeneds}) requires that
\be
g_k=-Res_{\infty+} (x^{-k} dS)~~,
\ee
which can be verified \cite{cekov}. 
Using (\ref{diffgeneds}) the meromorphic differentials $d\Omega_l$ can be 
written as
\bea
d\Omega_0&=&\frac{\partial dS}{\partial t_0}=\frac{g_{n+1} x^{n-1}}{y}dx+
\frac{1}{2}
\sum_{l=0}^{n-2}\frac{\partial f_l}{\partial t_0} \frac{x^{l}}{y}dx~~, \non\\
d\Omega_k&=&\frac{\partial dS}{\partial g_k}=\frac{W'(x)x^{k-1}}{y}dx+
\frac{1}{2} \sum_{l=0}^{n-2}\frac{\partial f_l}{\partial g_k} \frac{x^{l}}{y}dx
~,~~~k=1,\dots,n~~~,\non\\
d\Omega_{n+1}&=&\frac{\partial dS}{\partial
g_{n+1}}=\frac{[W'(x)x^{n}+ t_0 x^{n-1}]}{y}dx+\frac{1}{2}
\sum_{l=0}^{n-2}\frac{\partial f_l} {\partial g_{n+1}}
\frac{x^{l}}{y}dx~~.
\label{diffmero0}
\eea
In this framework, the prepotential $\mc{F}$ and so the special geometry can 
be introduced thanks to the Riemann bilinear relations which guarantee the 
integrability condition of the prepotential \cite{cekov}. We must define 
correctly the first derivatives of $\mc{F}$ with respect to both the periods 
and the coupling constants
\beq
\label{derprim}
\frac{\pa
\mc{F}}{\pa s_i}=\Pi_i=\frac{1}{2}\oint_{\beta_i} dS~~~,~~~
\frac{\pa \mc{F}}{\pa
t_0}=\Pi_0=\frac{1}{2}\int_{\infty-}^{\infty+}dS~~,\quad
\frac{\pa\mc{F}}{\pa g_k}=Res_{\infty+}\(\frac{x^{k}}{k}
dS\)~~.
\eeq
As we have seen in the previous section, the supersymmetry breaking 
contributions appearing in the
effective glueball superpotential (\ref{223}) are mixed second derivatives of 
the prepotential with respect to the Whitham parameters $g_k$. 
Starting from the expressions (\ref{derprim}) it results
\be
\label{dersecbis1}
\frac{\pa^2 \mc{F}}{\pa s_i \pa g_k}=
 Res_{\infty+}\(\frac{x^k}{k}d\omega_i\)~~,
\qquad
 \frac{\pa^2 \mc{F}}{\pa t_0 \pa g_k}=
Res_{\infty+}\(\frac{x^k}{k}d\Omega_0\)~~.
\ee
The right hand side of these formulae 
express the susy breaking contributions in (\ref{223}) as
geometrical quantities which can then be read directly as residues.
Nevertheless we have to
remind the interpretation of the mixed second 
derivatives
appearing in (\ref{223}). 
As already mentioned, they should be thought to come from an appropriate higher
degree system taking 
$g_j \to 0$ ($j>n+1$), with the genus of the curve
and $(t_0, s_i)$ kept fixed.
Using (\ref{dersecbis1}), the residues can be computed directly with $g_j=0$; 
therefore, $\frac{\pa^2 \mc{F}}{\pa t_0 \pa g_j}|_{g_j=0}$ and
$\frac{\pa^2 \mc{F}}{\pa s_i \pa g_j}|_{g_j=0}$ can be 
properly obtained from the curve of ($n+1$)--degree which depend
only on the couplings $g_k$, $k=1,\cdots,n+1$.
We can then 
extract all the mixed second derivatives, included those with respect to $g_j$,
using the $(n+1)$--degree geometry.

This 
simplification is one of the advantages of 
the embedding of the geometry in the Whitham framework.
With this approach we compute the holomorphic 
supersymmetry breaking terms in the effective glueball superpotential 
corresponding to a non--supersymmetric 
perturbation of the ($n+1$)--degree tree--level 
superpotential (\ref{Wtreedef}), without the explicit knowledge
of the prepotential $\mc{F}$.

\section{Tree--level cubic superpotential}
\label{results3}

We consider the simple case of a supersymmetric $U(N)$ gauge theory with 
tree--level superpotential
\be
\label{Wtreecubic}
W_{tree}(\Phi)=\frac{m}{2}\Tr\Phi^2+\frac{g}{3}\Tr\Phi^3~.
\ee
As suggested before we break supersymmetry promoting the coupling constants
of the tree--level superpotential to spurions (\ref{spurios1},\ref{spurios2}) 
deforming (\ref{Wtreecubic}) as
\be
\label{supalbdef}
W_{tree}(\Phi)=\frac{m+\theta^2 \Gamma_2}{2}\Tr\,\Phi^2+\frac{g+\theta^2 
\Gamma_3 }{3}\Tr\,\Phi^3+\theta^2\,\sum_{j>3}\frac{\Gamma_j}{j} \Tr\,\Phi^j~.
\ee
The geometry of the $\CN=1$ solution is described by the following complex
curve of genus one with meromorphic differential 
\bea
y^2&=&g^2(x-a_1)^2(x-a_2)^2+f_0+f_1 x~,
\label{curva3}\\
dS&=&y~dx=\sqrt{g^2(x-a_1)^2(x-a_2)^2+f_0+2 g t_0 x} \, dx~.
\label{dsnew5}
\eea
with $a_1=0$ and $a_2=-\frac{m}{g}$ the classical roots.

We do the computation of the supersymmetry breaking parts as a series
with small width of the cuts and then small values of $s_i$ and $t_0$. The 
approach is the same as in \cite{LargeN}.
In particular, we have considered the case of classical susy vacua with
unbroken gauge group and 
also the case with  $U(N)\to U(N_1)\times U(N_2)$ 
gauge symmetry breaking pattern.
Using (\ref{dersecbis1}), we compute directly the second mixed 
derivatives of the 
prepotential, i.e. the susy breaking contributions.
The details of the computations are in appendix A.
We express our results directly in terms of the physical glueball 
superfields $S_i~(i=1,\dots,n)$ using the change of variables 
(\ref{cambcoord}) at the end of the computation.
We will write explicitly only the novel supersymmetry breaking contributions to
the low--energy glueball superpotential referring the reader to the literature
\cite{LargeN,FujiOoko} for the well known $\CN=1$ part.\\

In the case $U(N)\to U(N)$, using (\ref{223}) with superpotential of the form 
(\ref{supalbdef}) 
we find
\bea
-\frac{1}{N}W_{eff}&=&
-\frac{1}{N}W_{eff}^{\CN=1}(S,m,g)~+\non\\
&&-~\theta^2 \Gamma_2
\[\frac{S}{m}\(~1+\sum_{k=1}^{+\infty}{\frac{3}{(k+1)!}}
{\frac{\Gamma(\frac{3k}{2})}{\Gamma(\frac{k}{2})}
{\(\frac{8g^2S}{m^3}\)^{k}}}~\)\]~+
\non\\\non
\eea
\bea
+~\theta^2\Gamma_3 \[~\frac{S}{g}\(~\sum_{k=1}^{+\infty}{\frac{2}{(k+1)!}}
{\frac{\Gamma(\frac{3k}{2})}{\Gamma(\frac{k}{2})}
{\(\frac{8g^2S}{m^3}\)^{k}}}~\)\] ~+
~~~~~~~~~~~~~~~~~~~~~~~~~~~~~~~~~~~~~
\non\\\non\\
+~\theta^2\Gamma_4  \[ \frac{m^4}{64g^4}\sum_{k=2}^{+\infty}\frac{1}{k!}
\( (k+1)\frac{\Gamma(\frac{1}{2}(3k-4))}{\Gamma(\frac{1}{2}k)}-
4 \frac{\Gamma(\frac{1}{2}(3k-1))}{\Gamma(\frac{1}{2}(k+1))}
  \)\(\frac{8 g^2 S}{m^3}\)^{k}\]~+~~\non
\\\non\\
~~~-~\theta^2\sum_{j>4}\frac{\Gamma_j}{j}\Bigg[ 
\frac{g}{j!}
\(\frac{\partial^j }{\pa \xi^j} 
\frac{1}{\sqrt{(g+m\,\xi)^2+f_0\,\xi^4-4gS\,\xi^3} } 
\)\Bigg|_{\xi=0}+\non ~\,~~~~~~~~~~~~\,~~~~~~~~~~\\
+\frac{m}{2(j-1)!}(1+Y)
\(\frac{\partial^{j-1}}{\pa \xi^{j-1}} 
\frac{1}{\sqrt{(g+m\,\xi)^2+f_0\,\xi^4-4gS\,\xi^3} } 
\)\Bigg|_{\xi=0}\,\Bigg]~,~~ 
\label{exactONEcut}
\eea
$Y$ and $f_0$ are functions of ($S,\,m,\,g$) whose expressions 
(\ref{Y(sigma)}, \ref{f0diS}) are given in appendix A. 

As a consistency check of our computation and focusing on the spurionic  
terms $\Gamma_2$ and $\Gamma_3$, we can compare the previous
result with the mixed second derivatives of the $\CN=1$ perturbative 
prepotential
\beq
{\mc{F}}=\frac{S^2}{2}\sum_{k=1}^{+\infty}{\frac{1}{(k+2)!}}
{\frac{\Gamma(\frac{3k}{2})}{\Gamma(\frac{k}{2}+1)}}
\(\frac{8g^2 S}{m^3}\)^k~~,
\eeq
derived for the first time in \cite{BIPZ} from the large-N matrix model. We 
find a complete agreement except 
the linear term ($\sim S$) in the series multiplied by $\Gamma_2$.\\
The appearance of the linear term can be explained in the following way. It is
known \cite{measure,DV,CDSW} that the measure in the matrix model partition 
function and also the allowed divergent modes on the complex curve
\cite{cachazovafa} give schematically a 
contribution like $(S-S\log(m \Lambda_0^2 / S))$ where
$\Lambda_0$ is a cut-off: this contribution together with the additive term 
($2 \pi i \tau S$) in the effective superpotential
gives the Veneziano--Yankielowicz superpotential \cite{Veneziano}.
The derivatives of this contribution w.r.t. the coupling $m$
give exactly the linear term appearing in (\ref{exactONEcut}) which
also agrees with what we have found using perturbative 
techniques (see Sec.6).\\
The supersymmetry breaking part coming from the quartic term (and also from the
higher ones) can be checked by comparison with the $\CN=1$ superpotential 
computed implicitly in \cite{ferrariElmetti} for a generic tree--level 
superpotential.
By evaluating the derivative where all the coupling constants except ($m,~g$) 
are set to zero, we find agreement with their computation for all the finite 
order explicitly given by them.\\

In the case 
$U(N)\to U(N_1) \times U(N_2)$, we consider only $\Gamma_2,~\Gamma_3,~\Gamma_4$
in (\ref{supalbdef}) as source of susy breaking. 
Then, using (\ref{223})  
we have
\bea
-W_{eff}&=&-W_{eff}^{\CN=1}(S_1,S_2,m,g)~+
~~~~~~~~~~~~~~~~~~~~~~~~~~~~~~~~~~\non
\eea
\bea
&+&\theta^2 \Gamma_2 \Big[
(2N_2-N_1)\frac{S_1}{m}+(2 N_1-N_2)\frac{S_2}{m}
+30 (N_1-N_2) \frac{g^2}{m^4} S_1 S_2+ \non \\
&&~~~+~
3(5 N_2-2 N_1)\frac{g^2}{m^4}S_1^2+3 (2N_2-5 N_1) \frac{g^2}{m^4} S_2^2+O(S^3)
\Big]+\non \\
&+&\theta^2 \Gamma_3 \Big[
-2 N_2\frac{S_1}{g}-2 N_1\frac{S_2}{g}+20 (N_2-N_1) \frac{g}{m^3}S_1 S_2+\non\\
&&~~~+~
2(2 N_1-5 N_2)\frac{g}{m^3}S_1^2+2(5N_1-2N_2)\frac{g}{m^3} S_2^2+O(S^3)
\Big]+ \non \\
&+&\theta^2 \Gamma_4 \Big[
2 N_2\frac{m}{g^2}S_1+(2N_1+N_2)\frac{m}{g^2}S_2+
\frac{6}{m^2}(2 N_1-3 N_2) S_1 S_2+\non \\
&&~~~+~ \frac{9}{2}(N_2-2N_1)\frac{S_2^2}{m^2}-
\frac{3}{2}(N_1-4 N_2)\frac{S_1^2}{m^2}+O(S^3)
\Big]~.\label{TWOcuts}
\eea
where we show terms up to the quadratic order in $S$; we give in Appendix A a 
sketch of the computation.

We can check also this case using the results 
of \cite{cachazovafa} for a quartic tree--level superpotential. 
Taking the derivatives of their results with respect to the coupling constants
and then making the appropriate limit ($S_3=0$ and $g_4 \to 0$) we get 
exactly our supersymmetry breaking contributions.\\
Observe that, also in this case, linear terms appear in the
supersymmetry breaking series multiplied by $\Gamma$'s. 
These can again be understood as coming 
from the Veneziano--Yankielowicz piece of the effective superpotential.
In fact, the scales $\Lambda_i$ associated with each unbroken gauge group 
sector $U(N_i)$ are functions of the coupling constants as a consequence of the
threshold matching \cite{Treshold}; by taking
derivatives w.r.t. the couplings we get exactly those linear 
contibutions appearing in (\ref{TWOcuts}).\\
Finally we note that, up to the quadratic order in $S\equiv S_1$, we can 
consistently get our first result (\ref{exactONEcut}) 
from the second one (\ref{TWOcuts}) simply by setting ($S_2=0$, $N_2=0$).

\section{Perturbative arguments}

In this section we exploit the perturbative approach \cite{DGLVZ} to 
discuss, from a field theoretical point of view, our use of the $\CN=1$ 
prepotential to study the low--energy glueball superpotential in the case with 
broken susy. We consider only the case of unbroken $U(N)$ gauge group 
and tree--level superpotential for the adjoint chiral superfields given by
$W(\F)=\sum_{k=2}^{n+1} \frac{G_k}{k}\Tr\,\F^k$ where $G_k= g_k+\te^2 \G_k$
are the spurionic coupling constants.

We recall that, because of holomorphicity, 
in the $\CN=1$ case the effective
superpotential is a function only of the coupling constants
$g_k$ and not of the $\Bar{g}_k$ \cite{Seiberg9309335,DGLVZ,CDSW}.
In our case susy is broken by the spurions and
holomorphicity in the couplings is not any longer a 
property of the superpotential.\\
In a perturbative framework the spurions $G_k$ can be thought as ordinary 
background chiral superfields.
We can then think susy unbroken and the perturbative computations in a 
superspace approach go using the usual $D$--algebra \cite{SUPERSPACE}.
The effective action will be schematically of the form
\beq
\int d^2\te d^2\teb\, {\cal K}(G_k,\Bar{G}_k,D^2G_k,\Db^2\Bar{G}_k,
\cdots,S,\Bar{S})
+\int d^2\te\, W_{eff}(G_k,S)+{\rm h.c.}~~,
\label{a0f}
\eeq
where the superpotential $W_{eff}(G_k,S)$ is constrained to be a holomorphic 
function\footnote{We are thinking 
about the case with masses in the Wilsonian approach for which the 
nonrenormalization argument works without IR patologies.} of $G_k$.

If we choose the particular supersymmetric configuration in which all the
chiral superfields $G_k$ are equal to the constants $g_k$, 
without any dependence on $\te_\a$, then 
the $\CN=1$ effective superpotential is $W_{eff}(g_k,S)$ and 
hence its holomorphicity \cite{Seiberg9309335}.

If we choose instead the 
configuration $G_k=g_k+ \te^2 \Gamma_k$ ($\Gamma_k\ne 0$),
we break susy and furthermore we 
will have two kind of contributions to the glueball superpotential.\\ 
The first ones come from $W_{eff}(g_k+\te^2\G_k,S)$ in (\ref{a0f})
and are the holomorphic ones we have studied in the previous sections. 
We call them the holomorphic contributions.\\
The others are $D$--terms contributions 
holomorphic in $S$ but not necessarily in the coupling 
constants $g_k$, $\Bar{g}_k$, $\Gamma_k$ and $\Bar{\Gamma}_k$
which come from particular contributions to ${\cal K}$ in (\ref{a0f})
and which can be written as $\int d^2\te$ integrals 
contributing to the glueball
superpotential\footnote{For example, 
the reader could think about two terms like 
(with $W^\a=i\Db^2(e^{-V}D^\a e^V)$ \cite{SUPERSPACE})
$\int d^2\te d^2\teb\,G(g,\Bar{g},\Gamma,\Bar{\Gamma},\te^2)
\teb^2\Bar{\Gamma}\,S^p= 
\int d^2\te\,G\, \Bar{\Gamma}S^p$ 
or
$\int d^2\te d^2\teb\,H(g,\Bar{g},\Gamma,\Bar{\Gamma},\te^2)
\Tr[i(e^{-V}D^\a e^V)W_\a]\Gamma S^q= \int d^2\te\,
H\,\Gamma S^{q+1}$ where $G$ and $H$ are functions of 
$g,\Bar{g},\Gamma,\Bar{\Gamma},\te^2$ and not $\teb^2$.}. These terms in the 
$\CN=1$ case ($\Gamma\to 0$) are zero.

Here we adopt a pragmatic 
attitude and we study only those 
contributions to the glueball superpotential which can be computed using the 
powerful perturbative techniques developed in \cite{DGLVZ} for the $\CN=1$ 
case.\\
In \cite{DGLVZ} the perturbative series was generated 
using only the propagator of the chiral matter superfield sector and the 
antichiral superfield $\Fib$ was integrated out. This was the central point 
for their simplifications.
In order to be able to integrate out $\Fib$ as in \cite{DGLVZ} we must have 
interactions only in terms of the chiral superfield $\F$ and then we consider
the following UV action
\bea
S(\F,\Fib)&=&\int d^4x\,d^4\te~\Tr~e^{-V}\Fib e^V\F-
\int d^4x\,d^2\te~{\frac{m}{2}}\Tr\,\F^2-
\int d^4x\,d^2\teb~{\frac{\mb}{2}}\Tr\,\Fib^2+\non\\
&+&\int d^4x\,d^2\te~{\frac{1}{2}}(\te^2\Gamma_2)\Tr\,\F^2+
\int d^4x\,d^2\te~\sum_{k=3}^m\frac{1}{k}(g_k+\te^2\Gamma_k)\Tr\,\F^k~~,
\label{StartPert}
\eea
where all the antiholomorphic interactions 
$\int d^2\teb\[{\frac{1}{2}}(\teb^2\Bar{\Gamma}_2)\Tr\,\Fib^2+
\sum_{k=3}^m\frac{1}{k}(\Bar{g}_k+\teb^2\Bar{\Gamma}_k)
\Tr\,\Fib^k\]$ are neglected. Furthermore,
since we are interested in the glueball superpotential it is also possible to
do the usual
simplifications of \cite{DGLVZ,CDSW} finding as the
relevant action\footnote{${\cal W}_\a=[W_\a,\cdots\}$ is 
the spinorial gauge field strength adapted to the action, as a 
graded--commutator, on the adjoint representation of the $U(N)$ gauge group.}
\beq
\int d^4x\,d^2\te~\Bigg{\{}{\frac{1}{2\mb}}\F[\Box-i{\cal W}^\a \pa_\a-m\mb]\F
+W^{int}_{tree}(\F)\Bigg{\}}~~,
\label{effS}
\eeq
where $W^{int}_{tree}$ in our susy broken case consists in the second line of
(\ref{StartPert}).
The difference with respect to \cite{DGLVZ} is that the tree--level
superpotential is now defined in terms of spurionic coupling constants.

Now, from (\ref{effS}), it is clear that the glueball
superpotential we are going to compute will be holomorphic in $S$ 
and in all the coupling constants except, at most, for the mass. 
In particular we observe (we refer to Appendix B for the details) 
that we can have contributions only 
of the following form
\beq
\int d^2\te \left\{ 
W_{eff}(G_k,S)
+{\frac{1}{\mb^2}}\sum_l {\cal B}_{l}(g_k,\G_k,\te)S^l
\right\}~~.
\label{GlueS}
\eeq
$W_{eff}(g_k+\te^2\G_k,S)$ is the holomorphic contribution we have already 
defined. Instead,
the second part of (\ref{GlueS}) is a particular subclass of the $D$--term 
contributions discussed before where
${\cal B}_{l}$ are holomorphic in all $g_k$, $\Gamma_k$ and possibly 
depend also on $\te^2$.

A careful perturbative analysis of (\ref{GlueS}) shows 
that all the coefficients ${\cal B}_{l}=0$ vanish $\forall l$ and that the
contributions to the glueball superpotential we are computing have the 
following form
\beq
\int d^2\te\[ N\te^2\G_2 {\frac{S}{m}}+ N{\frac{\pa {\cal F}_0}{\pa S}}\]
~~~~~~{\rm with}~~~~~~{\cal F}_0=\sum_l{\cal F}_{0,l}(g_k+\te^2\G_k)S^l~~.
\label{Wfattor}
\eeq
We refer the interested reader to Appendix B for the technical details of our 
perturbative computations.\\
In (\ref{Wfattor}) ${\cal F}_{0,l}$ are the planar amplitudes with $l$ index 
loops of the dual matrix model \cite{DV,DGLVZ} where the coupling
constants are in this case the spurions $G_k=g_k+\te^2\G_k$. The first term in
(\ref{Wfattor}) 
is given by a $1$--loop diagram with one vertex
$\frac{1}{2}\te^2\G_2\Tr\,\F^2$. 
This term is associated with the $1$--loop matter
contribution to the Wilsonian beta function for the gauge kinetic term which
is implicit in the nonperturbative Veneziano--Yankielowicz superpotential.
In the previous section we have seen that this term is also given by
the geometrical methods.

We conclude that, within our stringent approximations and in the case of 
unbroken $U(N)$, the effective glueball superpotential in the presence of 
spurions (\ref{Wfattor}) can still be deduced from the $\CN=1$ holomorphic 
superpotential supporting the results of the previous sections.

\section{Conclusions}

The Dijkgraaf--Vafa conjecture with supersymmetry breaking is the subject of 
this work.
We have considered the simple case of $U(N)$ gauge theory with massive 
adjoint chiral matter multiplet with a polynomial tree--level superpotential.
We have studied the case where supersymmetry is broken 
in the tree--level superpotential by promoting the coupling constants to 
chiral spurions.
We have considered their $F$--components as non--supersymmetric small 
perturbations of the $\CN=1$ gauge theory and we have discussed how
holomorphy can still play a role.
The non--supersymmetric holomorphic contributions to the effective low--energy
glueball superpotential have been derived with geometrical methods 
embedded in the Whitham framework as well as with techniques of
superfield formalism with spurionic fields.

Non--holomorphic $D$--terms, soft breaking via gaugino mass, low energy vacua
are open to investigation. This goes beyond the information encoded in the 
holomorphic matrix model that we have used so far.

\section*{Acknowledgements}

\noindent
We would like to thank M. T. Grisaru, S. Penati and A. Zaffaroni for 
useful comments and discussions. This work is partially supported by INFN and
MURST under contract 2003-023852-008 and by the European Community's Human
Potential Programme under MRTN-CT-2004-005104.

\newpage

\appendix

\section{Solution of the broken superpotential}
\label{appA}

In this appendix we show the main tools and details used for the computation
in the cubic tree--level superpotential of section \ref{results3}.

We have already written the genus one Riemann surface characterising the
solution (\ref{curva3}, \ref{dsnew5}). We observe that there is one
holomorphic differential $\frac{dx}{y}$ defined on this surface. This 
differential can be expanded around the point at infinity in powers of 
$\xi=1/x$
\be
\frac{dx}{y}=\sum_{k=0}^{\infty} R_k \xi^k d\xi~~, \qquad 
\label{errek}
R_k=\frac{-1}{k!} \frac{\partial^k}{\pa \xi^k} 
\(\frac{1}{\sqrt{(g+m~\xi)^2+f_0~\xi^4+2gt_0~\xi^3} }\)
\Bigg|_{\xi=0}~~,
\ee
where $R_m$ are functions of $g_k$, $t_0$, $f_0$ and can be simply computed by 
power expansion of $y$. The normalized holomorphic differential $d\omega$ is 
then
\be
\label{diffolo}
d\omega=\frac{1}{h_0} \frac{dx}{y}~~,
\ee
where we have introduced the following quantities\footnote{The $\alpha$--cycle 
encircle counterclokwise the second cut accordingly to our conventions.}
\be
h_m=\oint_{\alpha} \frac{x^m dx}{y}~.
\label{h0}
\ee
The meromorphic differentials $d\Omega_k$ are defined
by\footnote{We denote $g_2=m$ and $g_3=g$ of (\ref{Wtreecubic}).}
\be
d\Omega_0= \frac{\pa dS}{\pa t_0}=
g\frac{x dx}{y}+\frac{1}{2}\frac{\pa f_0}{\pa t_0}\frac{dx}{y}~,
\qquad \quad d\Omega_k= \frac{\pa dS}{\pa g_k} \qquad k=2,3~,
\ee
and are completely fixed by the normalization constraints
\be
\label{constdomega}
\oint_{\alpha} d\Omega_0 =0~~~,
\qquad \quad \oint_{\alpha} d\Omega_k =0 \qquad k=2,3~.
\ee
Then $d\Omega_0$ results to be
\be
\label{domega0}
d\Omega_0=\(g x - g \frac{h_1}{h_0}\)\frac{dx}{y}~,
\ee
where we used the first normalization condition of (\ref{constdomega}) 
which implies $\frac{\pa f_0}{\pa t_0}=-2g \frac{h_1}{h_0}$.
Collecting these formulas we can express the second derivatives of the 
prepotential characterizing the susy breaking terms in the effective 
superpotential (\ref{223}) for the case under consideration as follow
\bea
\frac{\pa^2 \mc{F}}{\pa t_0 \pa g_k}&=&
Res_{0}\(\frac{\xi^{-k}}{k}d\Omega_0\)=\frac{1}{k}\(g R_k-g \frac{h_1}{h_0}
R_{k-1}\)~,
\label{dersecmi3}\\
\frac{\pa^2 \mc{F}}{\pa s_2 \pa g_k}&=&Res_{0}\(\frac{\xi^{-k}}{k}d\omega\)=
\frac{R_{k-1}}{k}\frac{1}{h_0}~,
\label{dersecmi4}
\eea
where the $R_k$ are defined in (\ref{errek}).

In the case of unbroken gauge group ($U(N)\to U(N)$)
the cut associated with the $s_2$ variable
degenerate to a point with $s_2\to 0$ and the only variable is $t_0$.
The curve (\ref{curva3}) can be written as
\begin{equation}
\label{curvaONEcut}
y^2=g^2 (x-x_1)^2(x-x_3)(x-x_4)~,
\end{equation}
where $x_3,~ x_4$ are the extremal points of the 
first cut and $x_1$ is the
double zero of the curve where the 
second cut degenerates.
Following \cite{LargeN} it is useful to introduce the quantities
\begin{eqnarray}
&\Delta_{43}=\frac{1}{2}(x_4-x_3)~~~,~~~\Delta=(a_1-a_2)={\frac{m}{g}}~,\\
&Q=\frac{1}{2}(x_4+x_3+2 x_1)=(a_1+a_2)=-{\frac{m}{g}}~,\\
&I=\frac{1}{2}(x_4+x_3-2 x_1)
\label{cub12}
=\sqrt{\Delta^2-2 \Delta_{43}^2}~,\\
&x_1=\frac{Q-I}{2}~~~,~~~
\a={\frac{g^2}{m^3}}~~~,~~~\s=8\a S~.
\end{eqnarray}
The above relations can be proved comparing (\ref{curva3}) and
(\ref{curvaONEcut}). We have also directly introduced the physically relevant
variable
$S\equiv S_1=-\frac{t_0}{2}$.\\
Being interested in finding $\frac{\pa^2 \mc{F}}{\pa g_k \pa t_0}$ as in 
(\ref{dersecmi3}) we have evaluated $h_1/h_0$ in this case 
\beq
\label{h1suh0}
\frac{h_1}{h_0}=x_1={\frac{Q-I}{2}}~.
\eeq
Then we find
\begin{equation}
\label{444}
\frac{\partial^2 \mathcal{F}}{\partial t_0 \partial g_k}=
\frac{g}{k}\(R_k-\frac{h_1}{h_0}R_{k-1}\)
=\frac{g R_k}{k}+\frac{m R_{k-1}}{2 k}(1+Y)~,
\end{equation}
where $Y$ is defined as
\beq
Y={\frac{I}{\Delta}}=
\sqrt{1-\frac{2\Delta^2_{43}}{\Delta^2}}~.
\eeq
Then $I\equiv I(\Delta_{43}^2)$ is a function of $\Delta_{43}^2$.
Written in terms of $x_i$ the variable $t_0$ is
\bea
t_0=-Res_{\infty}(dS)&=&-Res_{\infty}[g(x-x_1) \sqrt{(x-x_3)(x-x_4)} \, dx]=
\non\\
&=&\frac{g}{16}(2 x_1-x_3-x_4)(x_4-x_3)^2=-\frac{g}{2}\Delta_{43}^2
~I
~.
\label{t0fin}
\eea
From (\ref{cub12}, \ref{t0fin}) we find
\beq
\s=(1-Y^2)Y~,
\label{eqs}
\eeq
which gives $Y$, and then $\frac{\pa^2 \mc{F}}{\pa g_k \pa t_0}$, as a function
of $\s=8\a S$.
Solving (\ref{eqs}) and taking the appropriate branch we obtain
\bea
Y
=\frac{2^{\frac{1}{3}}}{3\(\sqrt{\s^2-\frac{4}{27}}-\s\)^{\frac{1}{3}}}
+\frac{\(\sqrt{\s^2-\frac{4}{27}}-\s\)^{\frac{1}{3}}}{2^{\frac{1}{3}}}
=1-\frac{1}{2}\sum_{k=1}^{+\infty}
{\frac{(8\a S)^k}{k!}}
{\frac{\Gamma(\frac{1}{2}(3k-1))}{\Gamma(\frac{1}{2}(k+1))}}~.~
\label{Y(sigma)}
\eea
Once $R_k$ ($k=1,2,3$) are found from (\ref{errek}) we have the
first two softly broken terms of (\ref{exactONEcut})
\bea
\frac{\pa^2 \mc{F}}{\pa t_0 \pa m}&=&
-\frac{S}{m}\[~1+3\sum_{k=1}^{+\infty}{\frac{(8\a S)^{k}}{(k+1)!}}
{\frac{\Gamma(\frac{3k}{2})}{\Gamma(\frac{k}{2})}}\]~,
\label{dF2mt0}
\\\non\\
\frac{\pa^2 \mc{F}}{\pa t_0 \pa g}&=&
\frac{2S}{g}\sum_{k=1}^{+\infty}{\frac{(8\a S)^{k}}{(k+1)!}}
{\frac{\Gamma(\frac{3k}{2})}{\Gamma(\frac{k}{2})}}~.
\eea
For the third term of (\ref{exactONEcut}) the computation is a little
more involved. The derivative to be computed is
\be
\label{derpre4}
\frac{\partial^2 \mc{F}}{\partial t_0 \partial g_4}=\frac{g}{4} R_4+\frac{m}{8}
R_3 (1+Y)~.
\ee
The coefficients $R_3$ and $R_4$ are
\be
\label{r44}
R_4=\frac{f_0}{2 g^3}+\frac{6 m}{g^3} S -\frac{m^4}{g^5}~, \qquad
R_3=\frac{m^3}{g^4}-\frac{2 S}{g^2}~,
\ee
where the unknown function $f_0$ appears. To compute it we
integrate in $S$ the equation that can be obtained from the first constraint
in (\ref{constdomega}) and from (\ref{h1suh0})
\be
\frac{\pa f_0}{\pa t_0}=-2 g \frac{h_1}{h_0}=-2 g \frac{Q-I}{2}=m+ g I ~.
\ee
We then have
\be
\label{f0diS}
f_0[S]= -2 m S -2 g \int \frac{m}{g} Y[S]dS=
c_1-4 m S  +\frac{m^4}{8g^2} \sum_{j=2}^{+\infty}
\frac{(8 \alpha S)^{j}}{j!}
\frac{\Gamma(\frac{1}{2}(3j-4))}{\Gamma(\frac{1}{2}j)}~,
\ee
where $c_1$ is a function only of the couplings $m$ and $g$. Using the other
constraints in (\ref{constdomega}) that define the derivatives of $f_0$ with
respect to the couplings $(m,g)$ it can be proven that $c_1$ vanishes.
Finally, we use the formulas (\ref{derpre4}, \ref{r44}, \ref{f0diS})
with $c_1=0$ and find
\be
\frac{\partial^2 \mc{F}}{\partial t_0 \partial g_4}
=\frac{m^4}{64g^4} \sum_{k=2}^{+\infty}\frac{(8 \alpha S)^{k}}{k!} \Big(
(k+1)\frac{\Gamma(\frac{1}{2}(3k-4))}{\Gamma(\frac{1}{2}k)}-
4 \frac{\Gamma(\frac{1}{2}(3k-1))}{\Gamma(\frac{1}{2}(k+1))}
 \Big)~.
\ee
We observe that with these ingredients one has formally all the needed 
quantities to compute in a closed form, as power series of $S$, all 
the susy breaking terms in the effective superpotential (\ref{223})
coming from higher order supersymmetry breaking
deformation in the tree--level superpotential (\ref{Wtreedef}). 
These are functions only of $Y(S)$ and $f_0(S)$ as shown in (\ref{444}).
In fact the coefficient $R_k$ are defined as (\ref{errek})
and are functions only of $t_0=-2 S$, of the couplings, and 
of $f_0(S)$ which is known (\ref{f0diS}). At the end the result is 
(\ref{exactONEcut}).\\
 
The computation in the case of broken gauge group 
($U(N)\to U(N_1)\times U(N_2)$)
uses a procedure as \cite{LargeN}, making the calculation as a power series in
the width of the cuts. Our aim is again to compute the susy breaking terms 
appearing
in (\ref{223}) using the formulas (\ref{dersecmi3}, \ref{dersecmi4}).
The main difference with the unbroken gauge group case
is that now the curve does not degenerate
\be
y^2=g^2(x-x_1)(x-x_2)(x-x_3)(x-x_4)~~.
\ee
We introduce quantities analogous as before\footnote{Note that as concern the 
classical roots there are no modifications.}
\begin{eqnarray}
&\Delta_{43}=\frac{1}{2}(x_4-x_3)~~~,~~~\Delta_{21}=\frac{1}{2}(x_2-x_1)~~,\\
&Q=\frac{1}{2}(x_4+x_3+x_2+x_1)=(a_1+a_2)=-{\frac{m}{g}}~~,\\
&I=\frac{1}{2}(x_4+x_3-x_2- x_1)=\sqrt{\Delta^2-2 \Delta_{43}^2-2 
\Delta_{21}^2}~~.
\end{eqnarray}
We don't have anymore the simplification (\ref{h1suh0}) and we 
have to write the integrals $h_0$ and $h_1$ as power series in the widths of 
the cuts ($O(\Delta_{ab}^3)$).
We then find the inverse expression of the widths of the cuts $\D_{ab}$
as a functions of ($s_2,\,t_0$) and obtain ($h_0,\,h_1$) 
in terms of ($s_2,\,t_0$).\\
We have also to evaluate the parameters $R_k$. They have the form (\ref{errek})
but now $f_0$ has to be understood as a function of two variables and 
$t_0=-2 (S_1+S_2)$. Precisely
$f_0$ is a function of $t_0$ and $s_2$ which are the independent variables
and it is determined through the relations
\be
\frac{\partial f_0}{\partial t_0}=-2 g \frac{h_1}{h_0}~~~,~~~ 
\frac{\partial f_0}{\partial s_2}=\frac{1}{\frac{\pa s_2}{\pa f_0}}
=\frac{1}{\oint_{\alpha}\frac{dx}{2y}}=\frac{2}{h_0}~~.
\ee
The first equation comes from the normalization condition (\ref{constdomega}) 
while the second one is a consequence of the definition (\ref{SiPi1}) 
of the variable $s_2$ using
the explicit form (\ref{dsnew5}) of the differential $dS$.\\
We then compute directly the second derivatives of the 
prepotential, the susy breaking terms (\ref{223}), using 
(\ref{dersecmi3}, \ref{dersecmi4}).
At the end of the computation we change variables (\ref{cambcoord}) to express
the superpotential in terms of the physical glueball superfields $S_i$.
What we find is (\ref{TWOcuts}).

\section{Details on the perturbative approach}
\label{appB}

In this appendix we explore the details which give (\ref{Wfattor}) from 
(\ref{StartPert}). 
We will use and extend to spurions the method developed in 
\cite{DGLVZ,KrausShigemori} also reviewing, for reader convenience, 
their basic steps.

Starting from (\ref{StartPert}), the propagator is the same as in \cite{DGLVZ}
\beq
<\F\F>=\frac{-\mb}{ \Box-m\mb -i{\cal W}^\a \pa_\a}~~.
\label{propFF}
\eeq
The gauge field strength is considered constant then the bosonic
and fermionic integrations completely decouple in the computation \cite{DGLVZ}.
To compute contributions to the glueball superpotential, we will use the usual 
chiral--ring properties of $W_\a$ \cite{DGLVZ,CDSW}.
For example, we will use $\Tr (W_\a)^n=0$ with $n>2$.

Using the double line notation, a Riemann surface (oriented for $U(N)$) 
with genus $g$ is associated to each topologically relevant diagram with $L$ 
momentum loop and $l$ index loop, so that $L=l+2g-1$.
The $D$--algebra is exactly as in \cite{DGLVZ}. The only difference is
that performing the $D$--algebra some $\pa_\a$ can act on the $\te$ of the
spurions $G_k=g_k+\te^2\G_k$ giving new terms.\\
It is possible to do some general considerations using the constraints given
by the $D$--algebra structure, the
properties of $W_\a$ and the geometry of the diagrams in the amplitudes.\\
We fix a diagram with $L$ momentum loops and
$l=L-2g+1$ index loops. From the $W_\a$ properties it follows that, for a
relevant amplitude for the glueball superpotential, the maximal number
of allowed $W_\a$ is $2l$ otherwise we would have at least one index loop with
more than three $W_\a$.
Furthermore, in order to perform the fermionic loop integrations
it is necessary to have at least
$2L$ $\pa_\a$ and then at least $2L$ $W_\a$. The number of $W_\a$ ($\# W_\a$) 
in a non--trivially zero diagram then satisfies the inequality
\beq
2L\leq \# W_\a \leq 2l=2L+2-4g~~~.
\eeq
This implies that $g\equiv0$ and the only relevant diagrams
to be considered are planar. Moreover the relevant contributions to the
glueball superpotential have $\# W_\a=2L,2L+2$.

We consider first the case $\# W_\a= 2L$ i.e. $\# \pa_\a= 2L$.
In this case the $D$--algebra has to be done only inside the fermionic loops
and then no derivative acts on the background spurions. This is
equivalent to say that these contributions are insensible to the $\te$
dependence of the $G_k$. Then for these kind of terms we can reabsorb the
quadratic vertex $\frac{1}{2}\te^2\G_2\Tr\,\F^2$ into the propagator
(\ref{propFF}) by simply doing the redefinition $m\to m+\te^2\G_2$. This is 
clearly true exept a $1$--loop amplitude with one vertex
$\frac{1}{2}\te^2\G_2\Tr\,\F^2$ contracted with the propagator (\ref{propFF}) 
which gives the first term of (\ref{Wfattor}).
The resulting contribution to the glueball superpotential with $\# W_\a=2L$
is then given by (\ref{Wfattor}) and, except the linear term in $S$, these term
are computed perturbatively using the dual matrix model of the $\CN=1$ case.

Now, we consider the other case $\# W_\a=\# \pa_\a=2L+2$. All index loops are 
now saturated and we have a contribution proportional to $S^{(L+1)}$.\\
These contributions are nonholomorphic since they are proportional to
$\mb^{-2}$. In fact, the corresponding diagrams have a multiplicative factor
$\mb^{P_i}$ from the numerator of the $P_i$ propagators (\ref{propFF}).
Expanding the propagators (\ref{propFF}) at order
$({\cal W}^\a\pa_\a)^{(2L+2)}$, we have $P_f=P_i+2L+2$ bosonic propagators
${\frac{1}{p^2+m\mb}}$ expressed in momentum space.
By redefining the bosonic loop momentum variables $p^2\to\mb p^2$, from the
bosonic Jacobian we are left with a contribution $\mb^{2L}$ while from the
denominator of the bosonic propagators we have a term $\mb^{-(P_i+2L+2)}$.
Summarizing we have $[\mb^{P_i}][\mb^{2L}][\mb^{-(P_i+2L+2)}]=\mb^{-2}$.
Therefore, along the calculation we will set $\mb\equiv 1$ and multiply the
final result by $\mb^{-2}$.\\
In performing the calculation it is convenient to express the propagator
(\ref{propFF}) in the Schwinger variables
\beq
\int_0^{\infty}ds_i~\exp{[-s_i(p_i^2+i{\cal W}^\a_i\pa_\a+m)]}~.
\eeq
As in \cite{DGLVZ}, the bosonic contribution is given by
\beq
Z_{boson}=\frac{1}{(4\pi)^{2L}}\frac{1}{(\det M(s))^2}~~~,
~~~M_{ab}(s)\equiv\sum_is_iL_{ia}L_{ib}~~~,~~~p_i=\sum_a L_{ia}k_a~~~.
\label{Zbose}
\eeq
We note that $M(s)$ is an $L\times L$ matrix and then the denominator
of $Z_{boson}$ (\ref{Zbose})
is a homogeneous polynomial of degree $2L$ in $s_i$.
Furthermore, we have, from the fermionic integrations of these
diagrams, $2L+2$ $s_i{\cal W}^\a_i$ terms.
Then, at the numerator we have a homogeneous polynomial of degree $2L+2$ in
$s_i$. The degree in $s_i$ of the numerator results to be greater than the
denominator degree.
Thus, for the class of diagrams with
$\# W_\a=2L+2$ (certainly) there is no cancellation between the bosonic and
fermionic integrations in contrast with the case $\# W_\a=2L$ \cite{DGLVZ}.

Performing the $D$--algebra we realize that there are two distinct
possibilities depending on the way the two extra $\pa_\a$ are distributed
on the external spurionic terms. The first possibility is when two $\pa_\a$
act on one spurionic constant $\te^2 \G_k$.
This contribution would have a multiplicative factor
$[(\G_k)\prod_{v=1}^{(V-1)}(g_{k_v}+\te^2 \G_{k_v})]$
($V$ is the number of vertices of the considered diagram).\\
The second possibility is when the two $\pa_\a$ act on two different spurions.
In this case we have a multiplicative term
$[(\te^\a \G_k)(\te_\a \G_{k'})\prod_{v=1}^{(V-2)}(g_{k_v}+\te^2 \G_{k_v})]$.

Summarizing the previous considerations, the general structure of the glueball
superpotential, due to the integration of the matter fields considering only
the holomorphic part of the interaction vertices as in (\ref{StartPert}), is
\beq
\int d^2\te \ggl N\te^2\G_2{\frac{S}{m}}+
N \sum_l{\cal F}_{0,l}(g_k+\te^2\G_k)lS^{l-1}
+{\frac{1}{\mb^2}}\sum_l {\cal B}_{l}(g_k,\G_k,\te)S^l
\ggr~~~.
\label{glueS}
\eeq
The ${\cal B}_{l}$ are holomorphic in all $g_k$, $\Gamma_k$ and possibly 
depend also on $\te^2$.
${\cal B}_{l}$ are analytic in all the variables except $m$.
We will show that ${\cal B}_{l}=0$ $\forall l$ justifying
(\ref{Wfattor}).

To compute ${\cal B}_{l}$ we must perform the $D$--algebra
and treat the group theoretical factor. As in \cite{DGLVZ} to simplify the
fermionic integrations we can use the fermionic Fourier momentum
representation.
The novelty in the computation is due to the fact that now
there is also the $\te^2$ from the spurionic vertices to be Fourier
transformed. In particular, we have
\beq
\te^2=-\d^{(2)}(\te)=-\int d^2\pi e^{i\pi^\a\te_\a}~~~.
\eeq
We focus on a planar diagram with $L=l-1$ bosonic momentum loops with $P$
propagators and $V$ vertices. In particular we consider the case with only one
spurionic constant $\te^2\G_k$ on which the $D$--algebra acts nontrivially.
The other case is analogue.

The fermionic contribution results to be\footnote{The index $j_v$, depending on
$v=\{1,\cdots,V\}$, runs from $1$
to $k_v$ which is the degree of the interaction 
vertex $v$ as: $\Tr\,\F^{k_v}$.}
\bea
Z_{fermion}&=&
\int \prod_{v=1}^Vd^2\te_v\,\te_1^2 \prod_{i=1}^P\[e^{-s_i {\cal W}_i^\a
i\pa_\a}\d^{(2)}(\te_{v_i}-\te_{v'_i})\]
\non\\
&=&
-\int\prod_{i=1}^P d^2\pi_i d^2r\,\prod_{v=1}^Vd^2\te_v
\prod_{i=1}^P\[e^{-s_i {\cal W}_i^\a\pi_{i\a}}\]
e^{i(\sum_{j_{1}=1}^{k_{1}}\pi_{j_{1}}+r)^\a\te_{1\a}}
\prod_{v=2}^Ve^{i(\sum_{j_{v}=1}^{k_{v}}\pi_{j_{v}})^{\a}\te_{v\a}}
\non\\
&=&
-\int d^2\te\int \prod_{a=1}^{P-V+2} d^2\kappa_a
d^2r\,\d^{(2)}\Big(\sum_{j_{1}=1}^{k_1}\pi_{j_{1}}+r\Big)
\prod_{i=1}^P\[e^{-s_i {\cal W}_i^\a\pi_{i\a}}\]
e^{i(\sum_{j_{V}=1}^{k_V}\pi_{j_{V}})^\a\te_\a}
\non\\
&\Longrightarrow&
-\int d^2\te\int \prod_{a=1}^l d^2\kappa_a\,
\prod_{i=1}^P\[e^{-s_i {\cal W}_i^\a\pi_{i\a}}\]~~.\label{f2l}
\eea
In the second line $\pi_{j_{v}}$ are the spinorial momuntum connected to the
$v$--th $\te$--vertex and satisfy $\pi_{j_{v}}\equiv\pm\pi_i$ where the sign is
$+$ ($-$) if the spinorial momentum is going outside (inside) the vertex $v$.
In the last two line 
we exploit the relations
$\sum_{j_{v}=1}^{k_{v}}\pi_{j_{v}}\equiv 0$ ($v=2,\cdots,V-1$) with which
we define the remaining $l$ spinorial variables $\kappa_a$ from the independent
$\pi_i$. 
Furthermore, in the last line we have used the fact that we are searching for a
contribution to ${\cal B}_{l}$ which has $\# {\cal W}_\a=2l=2L+2$. Then, in the
expansion of $e^{-i(\sum_{j_{V}=1}^{k_{V}}\pi_{j_{V}})^\a\te_\a}$ we can keep 
only $1$, the term independent of $\te_\a$.
This is equivalent to say that only the term in which $\te^2$ of the spurion is
killed by two $\pa_\a$ contributes to ${\cal B}_{l}$.\\
At the end of the above manipulations we remain with $l$
fermionic integrations over the indipendent variables $\kappa_a$ and the 
$\pi_i$ are linear combinations of them
\beq
\pi_{i\a}\equiv\sum_{a=1}^l\widetilde{L}_{ia}\kappa_{a\a}~~.
\eeq
As in \cite{DGLVZ} we can implement the requirement of having two insertions
of ${\cal W}^\a$ for each index loop introducing $2l$ auxiliary grassmanian
variables ${\cal W}^\a_m$ adapted to the action on the adjoint representation
with
\beq
{\cal W}^\a_i\equiv\sum_{m=1}^l K_{im}{\cal W}^\a_m~~.
\eeq
The matrix $K$ is defined so that for each oriented $i$--th propagator the
$m$--th index loop can coincide and be parallel giving $K_{im}=1$; or coincide
and be anti--parallel giving $K_{im}=-1$; or not coincide giving $K_{im}=0$.

Summarazing we find from the fermionic integration
\bea
&&(16\pi^2S)^{l}\int\prod_{a,m=1}^l d^2\kappa_a d^2{\cal W}_m\
\exp{\left[-\sum_is_i\left(\sum_{a,m}{\cal W}_m^\a K^T_{mi}\widetilde{L}_{ia}
\kappa_{a\a}\right)\right]}\non\\
&&=(16\pi^2S)^{l}\int\prod_{a,m=1}^l d^2\kappa_a d^2{\cal W}_m\
\exp{\left[-\sum_{a,m}{\cal W}_m^\a \widetilde{N}(s)_{ma}
\kappa_{a\a}\right]}\non\\
&&=S^{l}(4\pi)^{2l}(\det \widetilde{N}(s))^2~~,
\eea
with
\beq
\widetilde{N}(s)_{ma}\equiv\sum_{i} s_iK^T_{mi}\widetilde{L}_{ia}~~.
\eeq
The relevant fact is that, for our class of diagrams which has an $S^2$
topology, the matrix $K$ has a nontrivial kernel.
In fact, for example, the vector $b_m$, whose components are all equal to one,
belong to the kernel of $K_{im}$\footnote{Our susy broken case is similar
to the situation which appears in the study of the perturbative reduction to 
matrix models for the case of $\CN=1$ supersymmetry and SU/SO/Sp(N) gauge 
groups developed in \cite{KrausShigemori} extending \cite{DGLVZ}.}.
This is simply due to the fact that in the case we are studying all momentum 
propagator lines have exact two index loop passing through them with opposite 
orientation; then, $\forall i$ there will be only one $K_{im'}=1$ and one 
$K_{im''}=-1$ ($m'\ne m''$) and $\sum_m K_{im}b_{m}=1-1=0$.
It follows that also the matrix
$[\widetilde{N}(s)]^T_{am}=\sum_{i} s_i\widetilde{L}^T_{ai}K_{im}$
has a nontrivial kernel indipendently of the explicit form of $\widetilde{L}$
which we have not analyzed in detail. Then $\det(\widetilde{N}(s))\equiv 0$.
This imply that ${\cal B}_{l}\equiv 0$ $\forall l$ as claimed before.

\end{document}